\documentstyle{article}     
\begin{document}
\begin{flushright}January $8^{th}$, 1997\\
1st revision: January $12^{th}$, 1997\\
2nd revision: January $26^{th}$, 1997\\
HEP-UK-0003
\end{flushright}

{\qquad } \vspace{2mm}
 \begin{center}
 {\Large Proposals on nonperturbative\\
superstring interactions
\quad\\ \quad\\ \quad\\
Lubo\v s Motl\\
\quad\\ \quad \\ 
{\it Faculty of Mathematics and Physics}\\
{\it at the Charles University in Prague,}\\
{\it Czech republic}\\
}
\end{center}
\vspace{3mm}

\def \eq#1{\begin{equation}#1\end{equation}}
\def \tb#1{\left(\begin{array}#1\end{array}\right)}
\def \abs#1{\left|#1\right|}
\def \bra#1{\left\langle #1\right\vert}
\def \ket#1{\left\vert #1\right\rangle}
\def \exp{\mbox{exp}}
\def \tp{\mbox{tp\,}}
\def \ignoruj#1{}

\begin{center}{\Large
\bf Abstract of this paper hep-th/9701025}\end{center}

We show a possibility that the matrix models recently
proposed to explain (almost) all the physics of M-theory may
include all the superstring theories that we know perturbatively.
The ``1st quantized'' physical system of one string seems to
be an exact consequence of M(atrix) theory with a proper
mechanism to mod out a symmetry.
The central point of the paper is the representation
of strings with $P^+/\varepsilon$ greater than one. I call
the mechanism ``screwing strings to matrices''. I also give the
first versions of the
proof of relation $R\approx \lambda^{2/3}$.
Multistring states are involved in a M(atrix) theory fashion,
replacing the 2nd quantization that I briefly review.
We shortly discuss the T-dualities, type I string theory and involving of
FP ghosts to all the system including the original one of Banks et al.

\vspace{3mm}
{\underbar {The first revision.}} I added a new proof of $R_1\approx
\lambda^{2/3}$ in pages 15-16 and some acknowledgements.

{\underbar {The second revision.}} Explanation of the origin of
level-matching conditions added.


\vspace{5mm}

\vspace{4mm}

\begin{tabular}{c}
\hline
E-mail: {\tt lmot2220@menza.mff.cuni.cz}\\
WWW-page: {\tt http://www.kolej.mff.cuni.cz/}
$\tilde{\,\,}${\tt lumo/e.htm}\\
\end{tabular}

\newpage

\section*{Contents}

\begin{enumerate}
\item{Review of the 2nd quantization and its death}
\item{M(atrix) theory in $0+0$ and $0+1$ dimensions}
\item{Compactification of one spatial coordinate}
\begin{itemize}
\item{A short note about the FP ghosts}
\end{itemize}

\item{Limits of the resulting IIA theory}

\begin{itemize}
 \item{The $E_8\times E_8$ heterotic string}
\end{itemize}
\item{Representation of the ``longer'' strings -- screwing them
to matrices}
 \begin{itemize}
  \item The origin of the level-matching condition
 \end{itemize}
\item{The scaling law $R\approx \lambda^{2/3}$}
\item{Compactification to nine dimensions}
\item{Getting type I strings from type IIB strings, open strings}
\item{Conclusions}
\end{enumerate}


\section{Review of the 2nd quantization and its death}

The second quantization has been for a long time the only
way to accomodate multi-particle states expressing naturally
that they are identical. The first candidate
to replace this machinery turned out to have form of matrix
models [1]. Different particles ``live'' in different blocks
of a block diagonal matrix and their permutation symmetry
(either Bose or Fermi) is contained in the gauge group of the
theory which is taken to be $U(N)$ in [1]. The
hamiltonian contains
squares of all the possible commutators thus for large distances
the physical states (with energy not too high) can be described
by almost commuting matrices which can be simultaneously
diagonalized. Only when the distances are small, the commutators
are not so important and the classical positions of particles
make no sense. This mechanism offers a natural realization
of the old idea that on the distances smaller than Planck
length the geometry does not work. But there are still many
puzzles concerning apparent increasing of the states' size
with $\varepsilon\to 0$.

\vspace{3mm}

The phrase ``the second quantization'' is usually used in
not so precise meaning -- as the
canonical quantization of a classical
field which has an infinite number of degrees of freedom,
for example the electromagnetic or the Dirac field although
the machinery of quantizing is the same as
the machinery for quantizing one classical particle.
Such a system has the same number of degrees of freedom
as a wave function in the one particle Schr\"odinger equation.

But we can show a real second quantization procedure. The first
quantization makes operators from classical observables like
$x,p$ and leads to a state vector (wave function). The second
quantization is based on the next step: we upgrade the values
of the wave functions in different points to be operators. Everything
can be written independently on a basis: the first quantized state
vector $\ket\psi$ is upgraded to an operator vector $\ket\Psi$
and the relation of orthonormality is upgraded to the
(anti)commutation relation
\eq{\left[\left\langle u\vert\Psi\right\rangle,
\left\langle
\Psi\vert v\right\rangle\right]_{grad}=\left\langle u
\vert v\right\rangle.}
The graded commutator is commutator or anticommutator according
to the grassmann parity of the states $\ket u$, $\ket v$.
The operator-ket-vector $\ket\Psi$ contains annihilation
operators $a_i=\left\langle i\vert\Psi\right\rangle$ while
the conjugate bra-vector $\bra\Psi$ contains the hermitean
conjugate i.e. creation operators $a^\dagger_i=\left\langle
\Psi\vert i\right\rangle$. Then we usually postulate a ground
state $\ket 0$ annihilated by whole $\ket\Psi$ and the excited
states are built by application of $\left\langle\Psi\vert
u\right\rangle$ creation operators.

Then we often write the main part of the second quantized hamiltonian
as the upgrade of the first quantized hamiltonian to multiparticle states.
For example, the first quantized operator $f$ is upgraded to the
second quantized $F$:
\eq{F=\bra \Psi f \ket\Psi}
Namely the identity operator $1$ is upgraded to the operator of number
of particles $N$. Graded commutators of the second quantized operators
are the upgrades of the corresponding first quantized ones. I will
not prove it here.
\eq{[\left\langle\Psi\vert f\vert\Psi\right\rangle,
\left\langle\Psi\vert g\vert\Psi\right\rangle
]_{grad}=\left\langle\Psi\vert[f,g]_{grad}\vert\Psi\right\rangle.}

Exactly this second quantization was used for strings in the light cone
gauge in works of Green and Schwarz and others. We built the canonical
second quantized hamiltonian from the one-string hamiltonian and
then we are adding those interaction terms expressing the locality
of splitting and joining strings (as well as crossing-over and others).
These terms are proportional to the coupling constant, a parameter
of the theory. Therefore the theory is perturbative. (I never
undestood if we need to add also next terms of higher orders in
$\lambda$ but let us not solve it here.)

Something from
this second quantization is necessary for the physics:
we need the states formed by two sufficiently distant
(and thus non-interacting) subsystems
to be a tensor product of these
subsystems. In this sense, there {\it must}
be a qualitative difference between a one-string state and
multistring states.

On the contrary, from an esthetical and intuitive point of view
there {\it should be not} a qualitative difference between
e.g. a state with two touching closed strings and a state
with one closed strings going along the same lines. Finally,
those states can be converted to each other by the interaction
terms. We probably cannot write the number of strings as some
sum (or integral) of a local quantity.

So we would like to get a formulation making no quantitative
differences between one- and two-string states. Of course, we must
keep the clustering property. Matrix models obey exactly
these requirements: the difference between states with
distinct number of strings is qualitative really only in the
free string limit and the clustering property is realized
in form of a block decomposition of matrices. The group of
permutations of particles is naturally contained in the
gauge group of these models.

\section{M(atrix) theory in $0+0$ and $0+1$ dimensions}

Let us begin in the almost beginning of the world -- in $0+0$
dimensions. We write the action as the dimensional reduction
of a $9+1$-dimensional Super-Yang-Mills theory to $0+0$
dimensions:
\eq{S=\mbox{Tr}\left(\frac 14{[X_\mu,X_\nu][X^\mu,X^\nu]}+
\theta^T\gamma_{0}\gamma_\mu[X^\mu,\theta]+
\beta\cdot 1\right)}
Here the $X$'s and $\theta$'s are hermitean matrices, $\theta$
is a real spinor of $spin(9,1)$ constrained to contain only
sixteen components of one chirality.
The $\beta$ term corresponds in some sense to $P^+$ (or
$P_{11}$ in the infinite momentum language):
its trace is the total $P^+$ proportional to the size of matrices
and the systems must be invariant under the group fixing
a quantity like
\eq{v^\dagger\cdot P^+\cdot v=
v^\dagger e^{iX^\dagger} P^+ e^{-iX}v.}
Therefore (for $P^+$ proportional to unit matrix) $X$'s must be
hermitean. The other choices of $P^+$ are nonequivalent
perhaps only if $P^+$ has a different signature. These
systems describing antibranes are also studied. The
symmetry group changes to a noncompact one and the
analytic continuation between $U(N)$ and $U(N-k,k)$
is responsible for the crossing symmetry [9].

Since the appearance
of $\gamma_{0}$ which must be included in the $32\times 32$
language (while in the $16\times 16$ language it can be
replaced by unit matrix) may look surprising, I will say
few words: it is necessary to make invariants in the same sense
as the Dirac conjugate spinor $\bar\psi=\psi^\dagger\gamma_0$.
The $32\times 32$ matrices $\gamma_\mu$ can be written using
the $16\times 16$ ones of $spin(9)$ denoted by $g_\mu$ as
\eq{\gamma_0=\tb{{rr}0&-1\\ 1&0},\quad
\gamma_{\mu=1..9}=\tb{{cc}0&g_\mu\\ g_\mu&0},\quad
\gamma_{chir}=\tb{{rr}1&0\\ 0&-1}}
All the matrices are real, $\gamma_0$ is antisymmetric
while the other nine are symmetric as is clear from the
following explicit form of those. These $16\times 16$
matrices can be written as these tensor products of
Pauli (or unit) matrices. Note the even number of
$y\equiv \sigma^2$ in each of them and that they
anticommute with each other:
\eq{g_{1\dots 9}=z111,x111,yz1y,yx1y,yyz1,yyx1,y1yz,y1yx,yyyy.}

Now we are on the lagrangian level (or better on the action level).
We have not a Hilbert space. In spite of that we can practice
in a limited sense the machinery described in [2] to obtain
the theory in $0+1$ dimensions described in [1].

We wish to mod out a continuous symmetry isomorphic to $R$
containing all the shifts of $X_0$:
\eq{X_0\mapsto X_0+\Delta X_0.}
According to [2] we must represent this group by a subgroup
of the gauge group. So we add two continuous indices
$t_m,t_n\in R$. So the matrices $X^\mu_{mn}$ are upgraded
to $X^\mu_{mn}(t_m,t_n)$. Here $t_m$ plays a similar role
as $m$ and $t_n$ as $n$. In other words, $t_m$ and $m$ together
form the left index while the $t_n$ and $n$ the right one.
So the matrices are tensored with operators on the space
of complex functions of a real variable.
Thus, for instance, the hermiticity condition takes the form
\eq{X^\mu_{mn}(t_m,t_n)^\dagger=X^\mu_{nm}(t_n,t_m).}
These indices are sufficient to represent the $\Delta X_0$
shift as the operator $\exp (i\Delta X_0\cdot t)$ which
has matrix components
\eq{Shift(t_m,t_n)=\delta(t_m-t_n)\exp(i\Delta X_0\cdot t).}
Note that if $t$ should be interpreted as a ``time'' then
the $\Delta X_0$ and thus also $X_0$ should be understood
as a dual quantity (``energy'').

Let us now write
the conditions of [2] for restricting the operators.
The $X_0$ shift has no influences to other matrices $Y$, so
they should obey ($Y$ is here understood as the operator
on the space of functions of $t$)
\eq{\exp(i\Delta X_0 t)Y\,\exp(-i\Delta X_0 t)=Y.}
It means that $Y$'s commute with all the waves and thus
with all the functions of $t$ -- therefore these matrices
are functions of $t$,
having matrix elements proportional to $\delta(t_m-t_n)$.
Therefore we can use one $t$ only.
For $X_0$ there is only one
modification -- the $X_0$ shift:
\eq{\exp(i\Delta X_0 t)X_0\,\exp(-i\Delta X_0 t)=X_0+\Delta X_0.}
Therefore $X_0$ has the form of a function of $t$ plus a derivative
according to $t$, creating the $\Delta X_0$ term.
\eq{X_0=x_0(t)+i\frac{\partial}{\partial t}.}
Therefore we have approached to the model of [1] where
the $t$-derivative is correctly generated and the function $x_0$
plays the role of the $A$ gauge potential in [1].
The trace in the $0+0$ model now includes also the
trace over the continuous indices $t_m,t_n$ and thus
the action will change to a $t$-integral.

Let us not forget the interesting point that the physical
time $t$ of our constructions is a {\it dual} variable
to the $X_0$ from the $0+0$ model.

\vspace{2mm}

Previous ideas were written mainly for amusing. Now we are standing
at the model of [1] with one time coordinate which already admits
a hamiltonian formulation, a Hilbert space and all these nice
things\dots

Nevertheless, we can learn many things already from the
construction above. By the way, the general idea that gauging
symmetry is achieved by looking at matrices which are gauge-equivalent
to their translations was first realized already by T.Banks
and his collaborators at least for the circular compactification.

\section{Compactification of one spatial coordinate}

Now we will compactify the coordinate $X^1$ to a circle
with period $R_1$. In other words, we will mod [2] the theory
in 0+1 dimensions by the group (isomorphic to $Z$) of all
the shifts by multiples of $R_1$:
\eq{X^1\mapsto X^1+k\cdot R_1,\quad k\in Z.}

We will translate the procedure of [6] directly to the
continuous basis. In [6] there is used the copying the
D0-branes to all the identified points, so there indices
$m,n\in Z$ are added to the matrices. We will make
directly the Fourier transformation of this procedure
and we will add two indices
$\sigma_1^m,\sigma_1^n$
parametrizing a circle in the same fashion as in the
previous section. The period of $\sigma_1^m,\sigma_1^n$
is taken to be $2\pi$.

In other words, we tensor the matrices with the space
of operators on the functions of an angle variable.

We represent the symmetry $X^1\mapsto X^1+k\cdot R_1$ by
the operator $\exp (ik\sigma_1)$. Note again that the
$\sigma_1$ is a momentum-like variable. (Its period should
be perhaps rather $\approx 1/R_1$.)

Now we can repeat the arguments of previous section and
show that all the operator-matrices except of $X^1$
are functions of $\sigma_1$. They contain $\delta(\sigma_1^m
-\sigma_1^n)$ again so we can use one $\sigma_1$.
And we can also show that the $X^1$ operator contains
a sum of a function of $\sigma_1$ and a derivative
according to $\sigma_1$ (understood to be multiplied
by the unit matrix):
\eq{X^1=x^1(\sigma_1)+iR_1\frac{\partial}{\partial \sigma_1}.}
This derivative acts on the other matrices, for instance
\eq{[X^1,X^2]=[x^1,x^2]+iR_1\frac{\partial x^2}{\partial
\sigma_1}.}
The hamiltonian had a form of trace and now the trace must
include also the trace (the integral) over the $\sigma_1$
variable. Therefore the hamiltonian has form like
\eq{H\approx \int_0^{2\pi}\!\!d\sigma_1\mbox{Tr}
\left(
\frac{\Pi_i(\sigma_1)\Pi_i(\sigma_1)}2
-\frac {[X^i(\sigma),X^j(\sigma)]^2}4+
\theta^T(\sigma_1)\gamma_i[X^i(\sigma_1),\theta(\sigma_1)]
\right),}
where the $X^1$ is understood to contain also $iR_1\partial/
\partial\sigma_1$.

We were dividing the trace by an infinite factor namely the
trace of unit matrix over the continuous $\sigma_1$.
(The same is true even for the first step which has given
us $0+1$ from the $0+0$ theory.)
But this
factor should be compensated by the same factor by which
we will rescale the transversal momentum $P^+$ (or $P_{11}$
in the infinite momentum frame formulation): the $P^+$
should be also proportional to that trace but let we
scale $P^+$ to be only a product of quantum and size $N$
of the matrices.

\subsection*{A short note about the FP ghosts}

By these procedures of compactification we obtain
Yang-Mills theories. The symmetry of the $0+0$ model
was $U(N)$. After we created a time, this symmetry
became local in time. It always has about the same number of
degrees
of freedom as one of the matrices, say $X^1$.

Now we can choose the $U(N)$ parameter to depend also
on $\sigma_1$. We could ask if the FP ghosts usual
in Yang-Mills theories should be included. Someone
could say that it is not natural to involve FP ghosts
which have no explanation in the process of the modding
out symmetries.

But I think that they have an explanation. Namely I would
like to argue that already the model [1] of Banks et al.
(and maybe even the model in $0+0$ dimensions)
should get ghosts among its fields. It is exactly what
is called in the chapter 3 of [3] (about the modern
covariant quantization) as a too complicated tool.

I think that
the proper mechanism to deal the FP ghosts is to
add them already to the model [1] in the canonical
form desribed for instance in [3], chapter 3 and
to require the states to have the zero ghost number
and to be annihilated by the BRST charge
\eq{Q\ket\psi=0}
and states $\ket\phi$ of the form
\eq{\ket\phi=Q\ket\alpha}
consider as trivial.
Then the modding out the symmetry [2] acts also on the
ghost terms in the hamiltonian and produces also
$\sigma_1$ derivatives from the commutators with $X^1$
present already in the original lagrangian (or
even hamiltonian?).

\section{Limits of the resulting IIA theory}

In the beginning I want to mention that I realized
that the derivation of IIA lagrangian from the M(atrix) theory
has been carried out already in the paper [1], maybe in all
aspects except of the representation of ``longer'' strings
combining both $\sigma$ and matrix-indices dependence and
quantitative relation between factors associated with free
string terms and the commutators, leading to the relation
$R\approx \lambda^{2/3}$. I appologize that I did not note it
immediately.

Let us now have a look to the hamiltonian we got
in the case of $N=1$. It is hard for me to call it
a ``D1-brane model'' or something like that since I
think that it is exactly the hamiltonian for
the IIA fundamental strings in Green-Schwarz formalism.
In the $N=1$ case all the commutators are zero and hamiltonian
becomes (16-component spinor $\theta$
of $spin(9)$ includes 8-component spinors of
$spin(8)$ of both chiralities $\gamma_1$)
\eq{H\approx \int_0^{2\pi}\!\!d\sigma_1
\left(
\frac{\Pi_i(\sigma_1)\Pi_i(\sigma_1)}2
+\frac{R_1^2}4\sum_{j=2}^9
\left(\frac{\partial X^j(\sigma)}{\partial\sigma_1}\right)^2+
iR_1\cdot\theta^T(\sigma_1)\gamma_1
\frac{\partial\theta(\sigma_1)}{\partial \sigma_1}
\right).}

In this section we just say that the limits
$R_1\to 0$ and $R_1\to \infty$ work quite well.
Let us have a look at the hamiltonian above.

In the case $R_1\to\infty$,
from the $[X^1+iR_1\cdot d/d\sigma,X^i]$
the most important term is the $\partial/ \partial
\sigma X^i$ term
which causes the $X$'s
(imagine classical matrices of numbers)
to be typically independent
on $\sigma_1$. We can replace $X^i(\sigma_1)$ by
$X^i$ and we are back to the original D0-branes model
of [1].

\vspace{2mm}

In the case $R_1\to 0$ the $\sigma_1$ derivative terms
are negligible. Then the typical configuration should have
the commutators sufficiently small because they are the
main contribution to the energy. So for every $\sigma_1$,
$X^i(\sigma)$, $i=2\dots 9$ can be approximated by commuting
matrices. But the basis in which they are simultaneously
diagonalizable can differ as we change $\sigma_1$
and the derivative of the basis is stored in the
gauge field $X^1(\sigma_1)$.
This change
can be fixed together with the gauge but there are some
$\sigma_1$-global effects, see the section ``screwing strings''.

\vspace{2mm}

The following paragraph is a part of text I consider wrong now.
Nevertheless, I have not deleted it.

\it

We discussed separately the cases with different $R_1$ but in
the next section we prove that these systems are equivalent. How
can this agree with present section? I think that the arguments
above are restricted to some quasiclassical configurations
which are typical for each $R_1$. States which are quasiclassical
in the system with hamiltonian having a given $R_1$ seem
perhaps highly nonclassically in the system having a very
different $R_1$.

\rm

\subsection*{The $E_8\times E_8$ heterotic string}

Before the wrong attempt
to prove the equivalence of systems of different
$R_1$, let us note very briefly
something about heterotic strings.
As in [2], the hamiltonian for IIA strings that we have found
has a symmetry transposing all the matrices, reversing the
sign of $X^1$ and $\Pi^1$
and multiplying spinors by $\gamma_1$ (that
we called $g_1$ a minute ago). Note that the combination of
the transposition and reversing the sign of $X^1$
keeps
its part $\partial / \partial \sigma_1$ invariant.
(The $\sigma_1$-derivative looks in the
terms of matrix elements as $\delta'(\sigma_1^m-\sigma_1^n)$.)
To be brief
(but maybe not
precise), let us represent this symmetry simply by
the unit matrix. In this case we restrict $X^{2\dots 9},
\Pi^{2\dots 9}$ as well as spinor components with
$\gamma^1=1$ to be symmetric real matrices,
while the $X^1$, $\Pi^1$ and the components of $\theta$
with $\gamma_1=-1$ to be antisymmetric real matrices.
Therefore (imagine $N=1$ case) these have no components
on the diagonal and we have only half of spinors after that.
The 32 real fermions to realize
the $E_8\times E_8$ (or $spin(32)
/Z_2$) symmetry must be included by hand. I think that
a correct way might be a vector representation of $SO(16)$
tensored with a vector of $O(N)$
(everything twice), the gauge symmetry
of the new model. The origin of these 32 fields is similar
as the existence of 32 D9-branes in type I theory
obtained by the modding out of $\sigma\to-\sigma$ symmetry
in the type IIB theory.

So this model could describe nonperturbative
heterotic strings, but I have no deeper explanation now.

\subsection*{Background independence on the $R_1$}

I realized that all this section is wrong and that there is
probably no background independence of the $R_1$. Hilbert
spaces
obtained by the compactification of M(atrix) theory probably
include only the states upon the particular values of moduli.
In spite of all that, I leave the wrong text here written in
italics: (just try it to correct)

\vspace{2mm}

\it
In this section I will show why there is only one
theory behind
IIA superstrings with different coupling constants
i.e. with different $R_1$. Let us remember that in
the perturbation theory the change of the coupling
constant can be reached by a vacuum condensate of
the dilaton field; the form of the dilaton's vertex
worldsheet operator has exactly the form to be able
to change the dimensionless coupling constant when
added to the lagrangian. Similar things are true
for other moduli -- fields/parameters desribing the vacuum.

The background independence was
being proved also in covariant string field theories [4].
In the ``pregeometrical'' formulation [5] these theories
have only the interaction term in the action and the
kinetic terms (and geometry) can be generated by a vacuum
condensate of the string field.

\vspace{2mm}

But let us return to our IIA hamiltonian.
$X^1$ contains $iR_1\partial/\partial\sigma_1$. We want
to change it without change of physics. In other words,
we would like to prove that there exist unitary operators
$U_{R\to R'}$ transforming one hamiltonian to the other
according to the formula
\eq{H(R'_1)=U_{(R\to R')}H(R_1)U^{-1}_{(R\to R')}.}
We can immediately translate this formula to a case
when $R_1$ and $R'_1$ differs only infinitesimally.
(A change of $R_1$ can be decomposed to a product
of infinitesimal changes.) In the infinitesimal language,
we need to find a hermitean operator $D$
\eq{iD(R_1)=\left.\frac{\partial U_{(R_1\to R'_1)}}{\partial R_1}
U^{-1}_{(R_1\to R'_1)}
\right\vert_{R_1=R'_1}}
such that the $R_1$ derivative can be transformed
into a commutator with $D$:
\eq{\frac{\partial H(R_1)}{\partial R_1}=
[iD(R_1),H(R_1)].}
Such an operator can be found. Note that the term
proportional to $R_1$ is always accompanied by $x^1$
since both are parts of $X^1(\sigma_1)$.

So although we have no variable $d$
directly dual to $R_1$ which could be used
as $D$, we can use the variable dual to $x^1$ which
always accompanies the $R_1$ proportional term.
So if we include $\mbox{Tr}\,\Pi_1$ into $D$,
it will give a correct commutator with all the $x^1$'s.
To get the derivative according to $R_1$ we must
add the $\partial / \partial \sigma_1$ operator to the
trace. So the total operator $D$ (it turns out
that it does not depend on $R_1$) could have a form like
\eq{iD=\int_0^{2\pi}\!\!d\sigma_1\cdot
\mbox{Tr}(\Pi_1)\cdot \frac{\partial}{\partial \sigma_1}
(\sigma_1)}
where the partial derivative denotes a standard density
of an operator shifting $\sigma_1$, having form of
\eq{\sum_Y Y'(\sigma)\cdot\frac{\delta}{\delta Y(\sigma)}}
where the functional derivatives are built from the dual
variables.

(By the way, I think that the rigid $\sigma_1$ shift should
be considered as a part of gauge symmetry and states should
be required to be invariant under these shifts, giving
string conditions like $N=\tilde N$.)

I hope that this construction can be generalized
to all the theories obtained by modding out
a group of symmetry operators with continuous parameters.
We have now showed that all the systems with different
$R_1$, or physically IIA strings with all possible
values of the coupling constant, are unitary equivalent.
They are equivalent even to theories where $R_1\to\infty$
but it need not to be completely true
to say that these theories
are exactly equivalent also to the $R_1=\infty$ case
i.e. to the model of [1]. But heuristically, there is only
one underlying theory.
\rm

\vspace{2mm}

Here the wrong (I think) text ends. The more precise calculations
of the commutators have not given the correct result. By the way,
if the hamiltonians for different $R_1$'s are not equivalent,
it is natural to consider $R_1$ as an observable
(this ensures that states upon all values of $R_1$
are included) which
has also its dual variable $D_1$ connected with zero-momentum
dilaton in some way. Since $D_1$ is not contained in the hamiltonian,
the hamiltonian commutes with $R_1$.

\section{Representation of the ``longer'' strings,\\
screwing strings to matrices}

In the beginning I want to say that in the Czech language there
is one word both for {\it matrix} and a {\it nut}: ``matice''.
This provides the motivation for the phrase ``screwing strings
to matrices'' (in Czech ``\v sroubov\'an\'i strun do matic'').
Maybe someone would prefer ``winding strings around matrices''.
But what is the idea?

\vspace{2mm}

If we ask how the multistring states are represented, we find a
usual answer in [1] with a natural modification: the matrices
$X^i(\sigma_1)$ whose matrix elements are the functions
$X^i_{mn}(\sigma_1)$ have a (block) diagonal form where
each block corresponds to one string. (The real physical
state is obtained from such an idealized one by the symmetrizing
over all the gauge group and other procedures.)

But now we have a new question: the transversal momentum $P^+$
in the light cone gauge
(or $P_{11}$ in the infinite momentum frame ideology)
is now naturally given again by the size of matrices $N$.
Note that in the light cone gauge superstring field theories
the total $p^+$ was always proportional to the length of
strings. Now the length is a multiple of a quantum. Which
multiple is given by $N$.

So the question is: how can we represent string with $P^+$
greater than the quantum of $P^+$ carried by the $N=1$ string?
For a time I thought that the right way is hidden in the
scaling of the $\sigma_1$ parameter and the strings with
a $k$-times
higher $P^+$'s are the elementary ones tensored with
$1_{k\times k}$. But the condition to allow the ordinary
interactions between strings were making the
representation more and more
complicated involving e.g. strings with period being
a $1/k$ part of the $\sigma_1$ period and so on.

\vspace{2mm}

Now I think that the string tensored with $1_{k\times k}$ matrix
is simply a set of $k$ strings because I found much more
convincing solution.

We have already said that in the $R_1\ll 1$ case the matrices
$X^i(\sigma)$, $i=2\dots 9$ should be simultaneously
diagonalizable but the basis in which they have all the
diagonal form can change with $\sigma_1$. This changing
with $\sigma_1$ is stored in $X^1(\sigma)$ which has a role
of a gauge field vector component.
Typically, because of the local $U(N)$ symmetry of the model,
the basis can be locally chosen to be independent of $\sigma_1$
but there can be global effects.

But we have a condition that the basis changes after
adding a period $2\pi$ to $\sigma_1$ again to a basis
where $X$'s are diagonalizable. But does it mean that
it must be a unit matrix? Are there other transformations
keeping the diagonal form?

Of course, there are: these transformations are the
{\it permutations of the eigenvectors}. Every permutation
can be decomposed to a product of cycles. And what
a cycle permuting $k$ eigenvectors denote? It denotes
simply a string with length ($P^+$ in the units
of its quantum) equal to $k$.

Because most of readers perhaps understand better to formulas,
I will write an equation of the screwing strings. Let $X^i$,
$i=2\dots 9$
denote the functions with periods $2\pi k$ expressing a string
with length $k$. What is the correct way to convert it into
simultaneously diagonalizable
matrices $X^i_{k\times k}(\sigma)$
of the size $k\times k$ with period $1\cdot 2\pi$?
\eq{X^i_{k\times k}(\sigma)=U(\sigma)
\cdot\mbox{diag}
(X^i(\sigma+2\pi),X^i(\sigma+4\pi),\dots,X^i(\sigma+2k\pi))\cdot
U^{-1}(\sigma)}
where the unitary matrix $U$ must obey
(written for the $k=4$ case to be clear)
\eq{U(\sigma+2\pi)=U(\sigma)
\tb{{cccc}
\circ&1&\circ&\circ\\
\circ&\circ&1&\circ\\
\circ&\circ&\circ&1\\
1&\circ&\circ&\circ\\}}
(Here $\circ$ denotes $0$.)
In the last matrix we can use any complex units
instead of $1$'s. Such a matrix function $U(\sigma)$
can be explicitly found for finite $k$ because $U(N)$
is connected. Later we will use a similar screwing for
the winding strings and there we will not be able
to find such functions since we will work with
a disconnected topological group. Note that $X^1$ must
keep the information about the changing of $U(\sigma)$
with $\sigma$ in a way like
\eq{X^1\propto i(\frac{\partial U(\sigma)}{\partial
\sigma})U^{-1}(\sigma)}

It is interesting to note that if we increase $k$
(and decrease the quantum of $P^+$) we transfer
the important information from the $\sigma_1$ dependence
to the dependence on the $U(N)$ indices. This change
is accompanied with a transfer of the
importance of the invariance under the
symmetry rigidly
shifting $\sigma_1$ (which is a part of the gauge
group) to the $U(N)$.

\subsection*{The origin of the level matching conditions}

The purpose of this subsection is to explain how the identities
like $L_0=\bar L_0$ between the left- and right- excitations
are contained in our understanding of the weakly coupled IIA
strings from the M(atrix) theory.

In the beginning I want to say one fact that can be confusing:
now everywhere we are talking about the IIA superstring theory
from M(atrix) theory -- about a corresponding model to [1]
but with one more spatial coordinate $X^1$ compactified!

Let us forget that the fundamental formulation of M-theory
arose from the study of D0-branes in the IIA theory and let
us see the importance of the model [1] in the $R,N/R\to
\infty$ limit where it describes M-theory in 11 dimensions.

It this paper, we describe a similar construction for IIA
theory itself which could be obtained by the study of
M-theory on $M^9\times T^2$.

\vspace{2mm}

But let us return to the main question of this subsection.
We want to show that only those
string states satisfying the level-matching
condition like $L_0=\bar L_0$ are allowed.
This condition expresses also an invariance under the
rigid shift of the $\sigma_1$ as we know from the
perturbative string theory.

Of course, the only condition for the states in the
present M(atrix) formalism is the invariance under the
gauge group. It means that all the conditions for states
must be elements of this group. We know that it works well
for the permutation of identical particles and in a future
paper we will show that also typical GSO projections are
simple elements of the gauge group.

In [1] the gauge group was simply $U(N)$. In our model, after
compactifying another coordinate $X^1$ to a circle with
period $R_1$, new ``coordinate-index'' $\sigma_1$
appears, theory becomes a kind of Yang-Mills
and the symmetry $U(N)$ now can depend also on $\sigma_1$.
The $U(N)$ has a diagonal subgroup $U(1)$ of the matrices
proportional to unit matrix and as far as I know, in [1]
all the states were identically invariant under this $U(1)$.
(I think so because the model [1] is a limit case of the
present construction where the $X$'s and $\theta$'s do
not depend on $\sigma_1$, so they are uncharged under
the $\sigma_1$ shifts which we will identify with
elements of $U(1)$ in a moment.)

Now the core of the simple proof: I will argue that the
rigid $\sigma_1$-shift\footnote{The period of $\sigma_1$
is taken to be $2\pi$.} by $\phi$
under which the states should be invariant
(it is the level-matching condition)
is expressed as the simple
(global, $\sigma_1$ independent)
gauge transformation
$e^{i\phi}\cdot 1_{(N\times N)}$.

The reason is very simple: in the process of the modding
symmetries, we identified a physical operation (namely
$X^1$ shift by $R_1$) with
the $\sigma_1$-dependent local $U(N)$ transformation
\eq{e^{i\sigma_1}\cdot 1_{(N\times N)}.}
Note that there is a special point $\sigma_1=0$ where this
phase equals one. Imagine that we would identify the
physical operation with
another element of the gauge group, namely
the product of $e^{i\phi}\cdot 1$
and the present $e^{i\sigma_1}\cdot 1$, i.e. with
\eq{e^{i(\phi+\sigma_1)}\cdot 1_{(N\times N)}.}
All the physics would be the same with only one
modification: there is $\phi+\sigma_1$ instead of
$\sigma_1$. So this variable is shifted by $\phi$
and this change just correspond to the $e^{i\phi}$
which we added to the product.

So the global $U(1)\subset U(N)$ transformation
$e^{i\phi}$ corresponds to the shift of $\sigma_1$ by
$\phi$. States must be invariant under whole local
$U(N)$ symmetry therefore also under this rigid shift.
So strings obey the level-matching condition.

Till now we have only found that the {\it total}
$\sum (L_0-\bar L_0)$ equals zero. Is it true also for
every single string in the block limit? The answer is
``yes'' because this condition for every single string
whose coordinates we write in a block is guaranteed
by the invariance under another element of the gauge group:
(in this example the string
whose level-matching condition we prove
lives in the $2\times 2$ block
on the left-top corner)
\eq{\tb{{cccc}e^{i\phi}&\circ&\circ&\circ\\
\circ&e^{i\phi}&\circ&\circ\\
\circ&\circ&1&\dots\\
\vdots&\vdots&\vdots&\ddots}.}

\section{Relation between compactification radius and the coupling constant}

Warning for the readers: the ideas below will not be too precise but
only give a chance that a correct proof of the relation [7] between
the compactification radius of M-theory and IIA (or $E_8\times
E_8$) coupling constant exists; in the following
pages I will give another proof whose idea should be more
reliable and works for both IIA and HE theories:
\eq{R_1\approx \lambda^{2/3}.}
All the formulas will be only schematical.
We will also use the sign $R$ instead of the previous
$R_1$.
Let us first write schematically
our form for the IIA hamiltonian:
\eq{H=Tr\int(\Pi^2+[X,X]^2+R^2X'^2+RX'[X^1,X]+
R\theta'\theta+\theta[X,\theta]).}
We want to study the weak coupling limit. So let us
first rescale the observables to achieve the
$R$-independent form for the free string terms
$\Pi^2,X'^2,\theta'\theta$. It is clear that we cannot
scale $\theta$ since the anticommutator with itself should
be independent of $R$ (proportional to one).
We can scale $\Pi=\Pi^{new}R^\alpha$ -- and thus
we must scale $X=X^{new}R^{-\alpha}$ to keep the commutator
of $\Pi,X$ constant. Then we must scale the whole
hamiltonian $H=H^{new}R^{2\alpha}$ to cancel the
powers of $R$ in the term $\Pi^2$.

We also want to cancel the powers of $R$ in the
$R^2X'^2$ term. Since $R^2X'^2=R^{2-2\alpha}X'^2_{new}$
should be a term of $H=H^{new}R^{2\alpha}$, the
$R$-cancellation requires
\eq{2-2\alpha=2\alpha\quad
\Rightarrow\quad \alpha=\frac 12.}
So if we rescale $\Pi=\Pi^{new}R^{1/2}$,
$X=X^{new}R^{-1/2}$ and $H=H^{new}R^1$, we see
that in formula for the new hamiltonian
$H^{new}=HR^{-1}$ this $R^{-1}$ cancels the
$R$ which was associated with $\theta'\theta$
so after rescaling also this term will be
automatically without $R$'s. Let us write the new
hamiltonian using the new $\Pi,X,\theta$
(all these fields below should have the index
$new$):
\eq{H=\mbox{Tr}\int
(\Pi^2+X'^2+\theta'\theta+
R^{-3}[X,X]^2+R^{-3/2}X'[X^1,X]+
R^{-3/2}\theta[X,\theta]).}
Remember that we would be happy to get the relation
$\lambda\approx R^{3/2}$ for the coupling constant.

Let us now discuss the new form of the hamiltonian:
the diagonal elements of $X^i,\Pi^i,\theta$,
$i=2\dots 9$ behave exactly
as in the free string light cone gauge theory.
Their typical sizes are independent of $R$ and so on.
The off-diagonal (complex) elements of these matrices
are constrained to be nearly zero: there is a harmonic
oscillator part of the hamiltonian for them
$p^2+R^{-3}X^2x_{off}^2$ where the potential comes from
the $[X,X]^2$ term.

The fermionic degrees of freedom have also the off-diagonal
elements which are also constrained by a similar condition.
I think that all the degrees of freedom in $\theta$ and
$X^i,\Pi^i$, $i=2\dots 9$ have the same character in one-string
and multistring states (see the section about screwing the
strings above). Even if it would be not the case, I think
that the factors for the radius-coupling relation would
cancel between these eight bosons and eight pairs of
fermions in the similar fashion as in the light cone
gauge IIA superstring field theories of Green and Schwarz.
(Of course, I admit that the main reason why I believe
this is the danger of getting $\lambda\approx R^{8k+l}$
which should be hardly equal to $R^{3/2}$ for $k\neq 0$.)

\vspace{2mm}

So the only field that I suspect from affecting the relation
is $X^1$. In fact, this is the field which stores the change
of the $U(N)$ gauge matrix. Let us consider the $U(2)$ case,
addmitting one-string and two-string states.

Let us realize that a $U(1)\times U(1)$ subgroup of the
full $U(2)$ local group is the group keeping the diagonal
form of $X^{2\dots 9}$ and $\Pi^{2\dots 9}$. So I think that
in a way this subgroup can be fixed and we can put the
diagonal elements of $X^1$ to zero (I am talking in a basis
where $X^i(\sigma)$ for given $\sigma$ are diagonal).

The off-diagonal (complex) element of $X^1$ (the element
accross the diagonal
is just its hermitean conjugate) is the field responsible
for the interactions. Note that in the mechanism of screwing
strings, we can keep $U(\sigma)$ everywhere
(for all $\sigma$'s) as the unit
matrix except of the interaction point. Here if we want to
create a one-string state from a two-string state, we
must make the transformation corresponding to the permutation
of two eigenvectors in a close neighbouring of this
value of $\sigma$ where the interaction occurs.
Such a quick continuous transformation gives a nonzero value
to the off-diagonal elements of $X^1$, storing the
\eq{iU'(\sigma)U^{-1}(\sigma)}

Let us also think that the interaction occurs in one point only
so we will not write any other $\sigma$ dependence. For the
$X^1$ off-diagonal elements we have again a harmonic oscillator
hamiltonian with the potential proportional to $R^{-3}$.
This element is complex, we will talk only about its hermitean
part $x$ and at the end we will not forget to square the result
since the factor coming from the imaginary part is the same.
The hamiltonian looks like ($k,k'$ and so on does not depend
on $R$)
\eq{h=p^2+kR^{-3}x^2}
and the wave functions normalized to a constant
independent of $R$ contain the factor
\eq{R^{-3/4}\exp(-x^2k'R^{-3/2}/2).}
Our complete hamiltonian $N=2$ admits one-string states and
two-string states. Since for $R\neq 0$ the hamiltonian is not
free, it can transfer these states to each other. I think
that the correct factor of the corresponding ``interaction
term'', which is responsible for these transfers in
the 2nd quantized theory, is proportional to a scalar product
of a typical one- and two-string states.

These states for the corresponding configuration of strings
differ in the $X^1_{off}$ dependence. Let us write these dependences
for $1,2$-string states as (we omit $k$'s and numeric constants)
\eq{R^{-3/4}\exp(-R^{-3/2}(x-x_{1,2})^2).}
Their scalar product
(integral from the product over $x$)
is proportional to (no factor before the exponential
can arise because for $x_1=x_2$ we must get again the
normalization of the functions)
\eq{\exp(-R^{-3/2}(x_1-x_2)^2).}
For $R\to 0$, this function can be (up to a constant)
understood as the delta function:
\eq{R^{3/4}\delta(x_1-x_2).}
Here $x$ was considered as the hermitean part of the
off-diagonal element and the same factor comes from the
imaginary part. So the whole scalar product is proportional
to
\eq{\lambda\propto R^{3/2}}
as we wanted to prove.

\subsection*{Another proof}

This subsection was added before the first revision of the paper.
I tried to generalize the proof above also to the HE theory
but it was giving a different exponent due to the fact that
the unitary group is replaced by the orthogonal (or symplectic)
and the field responsible for interactions is not already
just the complex off-diagonal element of $X^1$. Now I offer
you probably even simpler proof which gives correct results
for both IIA and HE theories.

In the following ideas I will use another rescaling which
could be called a ``$\sigma$-locally $R_1$-independent''
redefinition of fields. I mean that all the $R_1$ dependence
will be stored in the period of $\sigma_1$ which we will
rescale now, too.

Let us write the hamiltonian of IIA strings schematically
again:
\eq{H=\mbox{Tr}\int d\sigma_1
(\Pi^2+[X+iR_1\frac\partial{\partial \sigma_1},
X]^2
+\theta[X,\theta]).}
Now we will rescale also $\sigma_1$. The redefinitions are:
\eq{\sigma_1=\sigma^{new}R^{-\kappa},
\,\,\Pi=\Pi^{new}R^{\kappa-\lambda},
\,\,X=X^{new}R^\lambda,
\,\,\theta=\theta^{new}R^{\kappa/2}.}
The appearance of $\kappa$ and $\kappa/2$ in
$\Pi$ and $\theta$ respectively where $\kappa$
is the exponent for rescaling $\sigma_1$ is a
consequence of the need to keep the commutator
of $X,\Pi$ and the anticommutator $\theta$ with
itself proportional to $\delta(\sigma_1-\sigma'_1)$
where $\sigma_1$ is rescaled. Furthermore, we
add the rescaling of $X$ by $R^\lambda$
(and the inverse one for $\Pi$ to keep their
commutator invariant).

Now we wish the hamiltonian written in terms of new fields
to be independent of $R_1$. The hamiltonian becomes (all the
quantities below should have index $new$)
\eq{H=R_1^{-\kappa}\mbox{Tr}\int d\sigma_1
(R_1^{2\kappa-
2\lambda}\Pi^2+
[R_1^\lambda X+iR_1^{1+\kappa}
\frac\partial{\partial \sigma_1},R_1^\lambda X]^2
+R_1^{\lambda+\kappa}\theta[X,\theta]).}
We can also rescale whole hamiltonian so the
conditions for $R_1$ independence are that all the
terms inside the bracket have the same power of $R_1$.
The $R_1$-free summing of $X_1$ and the derivative gives
\eq{\lambda=1+\kappa.}
The same factor in the $\Pi^2$ term and in the $[X,X]^2$ term
gives the first condition below while the equality of factors
associated with $\theta[X,\theta]$ and $[X,X]^2$ gives the
second condition below
\eq{2\kappa-2\lambda=4\lambda,\quad \lambda+\kappa=4\lambda.}
Fortunately, these two conditions are equivalent thus we
can solve these three equations and we obtain
\eq{\kappa=3\lambda \,\Rightarrow\,
\lambda=1+3\lambda \,\Rightarrow\, \lambda=-\frac 12,\,
\kappa=-\frac 32.}
Thus the period of the new $\sigma_1^{new}$ where the hamiltonian
is locally independent of $R_1$ is $2\pi R_1^{-3/2}$
if the period of the old $\sigma_1$ was $2\pi$.

We have again got this magical power of $R_1$. How can we
prove that the interactions are in the first approximation
proportional to $R_1^{3/2}$? I would argue following way:

Imagine that $N$ is large (the size of matrices).
If we search the states which are the most close
to the same free states (states in $R_1=0$ limit),
the correct way after the $R_1^{-3/2}$ times
enlarging the period of $\sigma_1$ should be that
we use the same functions which must be
$R_1^{-3/2}$ times less screwed to the matrices
(see the previous section). Then let us have a look on a point
of a string which is waiting to interact with another
string and make the ``crosing-over'' operation. How
large are its chances to do that?

The interaction must occur locally in $\sigma_1$ so
our point can make the interaction only with points
from other (or the same) string having the same $\sigma_1$.
But if there are $R_1^{-3/2}$ times less such points
(since the string is screwed into the matrix
$R_1^{-3/2}$ times less) then also its chances to make
the interaction is $R_1^{-3/2}$ times smaller, so the
strength of interaction is in the first approximation
proportional to
\eq{\lambda\approx R_1^{3/2}.}
This argument should be valid both for IIA and HE
theories.

\section{Compactification to nine dimensions}

I was not able to obtain a similar non-perturbative
formulation of the IIB superstring theory. The main reason
is that I cannot imagine what a counterpart
of the terms $\theta^T\gamma_i[X^i,\theta]$
could look like in the case of IIB strings
where only components with $\gamma_1=1$ are present because
all the $\gamma_i$ matrices (except of $\gamma_1$)
map the $\gamma_1=1$ spinors to $\gamma_1=-1$ spinors
and vice versa.

Nevertheless, it is easy to imagine how the compactification
of the second coordinate $X^2$
to a circle of period $R_2$
should look like. We must add
$\sigma_2$ such that $X^2$ plays a role of the gauge
field component. For $R\to\infty$ we reproduce again
the original theory since the $\sigma_2$ dependence
looses its importance.

For $R_2\to 0$ we should get IIB theory but I repeat that
I cannot imagine even the terms of its hamiltonian in our
formulation which it should contain.

Let us briefly say something about the winding strings
around the compactified $X^2$ coordinate. In the process
of $X^2$ compactification, we identify the physical
operator $\exp(iR_2\Pi_2)$ with $\exp(i\sigma_2)$.
We can do a similar trick as in the explanation of
the longer strings using their screwing.
The unitary matrix after the $2\pi$ shift of $\sigma_1$
which equaled in the latter case to the permutation
matrix can now be equal to $\exp(i\sigma_2)$
which we identified with the $X^2$-shift
$\exp(iR_2\Pi_2)$ -- so we must add twisted
sectors in the same fashion as in superstring theories [2].
These sectors will contain the winded strings around $X^2$.
The main difference from the permutation matrix
is the fact that this matrix being a function of $\sigma$
cannot be written explicitly (if we wish to
keep the matrices to be continuous functions
of $\sigma$'s), because the group
\eq{U(1)^{S^1}}
is disconnected. Maybe it sounds incomprehensible:
so I add following explanation:

The operator $\exp(i\sigma_2)$ is a part of the $U(N)$ local
gauge symmetry of our model. It corresponds to choosing
the $U(N)$ matrix as $\exp(i\sigma_2)$ times the unit matrix
from $U(N)$. But the factor before the unit matrix
can depend on $\sigma_1,\sigma_2$ -- for every pair of
$\sigma_1,\sigma_2$ we can choose one complex unit.
The group of all continuous functions from the interval
$(0,2\pi)$
(mapping the $\sigma_2$)
to the circle of complex units
\eq{\sigma_2\mapsto \alpha(\sigma_2),\quad \abs{\alpha(\sigma_2)}
=1}
is disconnected.
The group of components is isomorphic to $Z$.
The representatives of the components can be chosen to be
$\exp(ik\sigma_2)$, $k\in Z$. We must add sectors also from
the other components of the group.

If there is a similar formulation of IIB strings, we could
also ask how their T-duality is built in our formalism.
T-duality should exchange winding and momentum modes
which are given (I guess) as
\eq{\int\int \mbox{Tr\,}\Pi_2,
\quad \int\int\mbox{Tr\,}F_{12},}
where the integral of $F_{12}$ plays the role of the
gauge invariant generalization of
the winding number
$\int d\sigma_1(X'_2)$.
In other words, it should be possible to obtain a formulation
of IIB strings from M(atrix) theory compactified to 2-torus
with a small area. (I had to note the law $L_B=A_M^{-3/4}$
in Schwarz's lectures [10].) In the ``$\sigma$-local
$R$-independent'' formulation the uncompactified 10D limit
of IIB strings correspond to the infinite $\sigma_{1,2}$-torus
where only $spin(7)$ is manifest while from the two compactified
dimensions of M-theory we obtain only one of IIB strings
with its momentum density given by $F_{12}$. Such a formulation
would have manifest $SL(2,Z)$ S-duality and work is in progress.

\section{Getting type I strings from type IIB strings, open strings}

Let us return to ten dimensions.

In spite of that we found no particular form,
let us imagine that we have a IIB superstring
theory in our formulation. Such a theory should have a similar
symmetry $\sigma_1\to-\sigma_1$ (which we must combine with
$x^1\to-x^1$)
as in the previous formulation of superstrings.
This symmetry which we will call $S$ ($S^2=1$)
need not to be combined with any
transpositions like in [2].
This symmetry also exchanges left-going and right-going fermions.
Let us use the standard machinery
of [2] to obtain twisted sectors.
The operator $S$ in some sense restricts the
$(0,2\pi)$
(or $(-\pi,\pi)$)
circle of $\sigma_1$ to the line interval
$(0,\pi)$ only, but let us make the things more precise.

We identify $S$ with some matrix $S'$ in the $U(N)$ group.
(I will work with $N=2$.)
In this case all three Pauli matrices should be equivalent
because no transposition plays the game.
If we choose $S'=\sigma^3$, the $X$ matrices are restricted
to obey
\eq{\tb{{cc}X_{11}(\sigma)=X_{11}(-\sigma),
X_{22}(\sigma)=X_{22}(-\sigma),\\
X_{12}(\sigma)=-X_{12}(-\sigma),
X_{21}(\sigma)=-X_{21}(-\sigma).}}
From these equations it is quite clear that a classical
solution for our hamiltonian in $R_1\to 0$ limit contains
two independent open strings whose coordinates
are stored in $X_{11}(\sigma)$ and $X_{22}(\sigma)$,
respectively. The matrix is again quasi-diagonal
(diagonal in the limit $R_1\to 0$).
These strings are really open because of the conditions
for the functions to be even. Since $S$ also exchanges
left-moving fermions with the right-moving, we have also
the condition $\theta_L(0)=\theta_R(0)$. Thus our
$N=2$ system can describe correctly two open strings
of unit length (we mean the $\sigma_1$-interval
of length $\pi$).

Now let us use an equivalent choice $S'=\sigma^1$. By this
choice another class of solutions is visible because of the
conditions we obtain (twice):
\eq{X_{11}(\sigma)=X_{22}(-\sigma),
X_{21}(\sigma)=X_{12}(-\sigma).}
Here we see that the matrix elements are not constrained
to be neither even nor odd functions of $\sigma$.
We see another solution which again puts the
off-diagonal elements $X_{12},X_{21}$ to be zero
and $X_{22}(-\sigma)$ can be expressed using $X_{11}(\sigma)$.
The only condition for this remaining $X_{11}(\sigma)$
is $2\pi$-periodicity so we got one closed string of length equal
to two.

By combining these operations also with the permutation-screwing
of strings we can also get one open string of length equal to two.
In a similar way also open and closed strings of any integer
length are contained in our formulation. The strings are unoriented.
For closed strings it is clear since there is no priviledged
direction on them but open strings are unoriented as well.

Interaction of the ``crossing-over'' type are built in
this construction in the same manner as in the IIA theory
of closed strings only which we were discussing in previous
sections.

The interaction of the type ``joining of two open strings''
and the opposite ones can occure only in the points
$\sigma_1=0,\pi$.

\vspace{2mm}

Now we can also shortly discuss the appearance of the $SO(32)$
group. I think that the right solution is in putting
16 fermions in the fundamental representation of the
gauge group
to both ends of the $\sigma_1$ line interval
$(0,\pi)$. So one end is responsible for one $SO(16)$
and the other end for the other $SO(16)$ in the
$SO(16)\times SO(16)$ subgroup of $SO(32)$.
I do not understand the mechanism why just one fermion
should be excited in the points corresponding to an
end of open string while none in the other points.

We can also note that the $N=1$ case would always
contain one open string connecting one end $1\dots 16$
with the other $17\dots 32$. This idea can be generalized
to a formula that for odd
$N$ (the size of the matrices) we should get an odd
number of such $1\dots 16$--$17\dots 32$ strings.
Just such strings are even under the transformation of
$SO(32)$ reversing signs of 16 components while
keeping the other 16 invariant.
(The matrix $-1_{32\times 32}$ is the identity operator
because of the even number of strings' ends in all the
states.)

This operation (having the eigenvalues $\pm 1$)
is thus identified with $(-1)^N$ which is also
equal to
\eq{(-1)^N=\exp(\pi i N)=\exp(\pi i P_{11}/\epsilon)}
the shift of $X^{11}$ by a half period.

\section{Conclusions}

In this paper I showed that the type IIA string theory is a direct
consequence of the M(atrix) theory [1] and a proper mechanism of
compactification [2]. The one-string Hilbert space turns out to be
an exact copy of that of the first quantized approach. Multistring
states are contained in the M(atrix) fashion as block diagonal
matrices. On the contrary, strings with the transversal momentum
greater than one quantum are represented by a funny mechanism
called ``screwing strings to matrices''.

Now I think that the possibility of background independence
turned out to be wrong. So for a given $R_1$, we restrict
our reasoning to states upon a vacuum with a given coupling
constant. I have showed a preliminary version of the proof
of the expected relation $R\approx \lambda^{2/3}$ and also
explained why only sets of string states satisfying the
level-matching condition are physical.

Nevertheless, there are very many questions unsolved.
Is there a covariant version of the M(atrix) theory
allowing e.g. RNS formalism, non-flat background geometries,
eventually also a non-zero cosmological constant?
Will be able to obtain realistic models? Let us be
patient\dots

\newpage
\section*{References}
\begin{enumerate}
 \item T.Banks, W.Fischler, S.H.Shenker, L.Susskind:
 {\it M Theory As A Matrix Model: A Conjecture,} hep-th/9610043
 \item L.Motl:
 {\it Quaternions and M(atrix) theory in spaces with boundaries,}
\newline hep-th/9612198 (2nd version)
 \item M.B.Green, J.H.Schwarz, E.Witten:
 {\it Superstring theory,} 2 volumes, Cambridge University Press 1987
 \item Hata, Itoh, Kugo, Kunitomo, Ogawa:
 {\it Manifestly covariant field theory of interacting strings I,II,}
Physics letters B172, p.186, 1986
 \item Hata, Itoh, Kugo, Kunitomo, Ogawa:
 {\it Pregeometrical string field theory: creation of space-time
and motion}, Physics letters B175, p.138, 1986

 \item W.Taylor:
 {\it D-brane field theory on compact spaces,} hep-th/9611042
 \item P.Ho\v rava, E.Witten:
 {\it Heterotic and Type I String Dynamics from Eleven Dimensions,}
hep-th/9510209
 \item D.Kutasov, E.Martinec: {\it New Principles
for String/Membrane Unification,} hep-th/9602049
 \item V.Periwal: {\it Antibranes and crossing symmetry,}
 hep-th/9612215
 \item J.H.Schwarz: {\it Lectures on Superstring and M
Theory Dualities,} \newline hep-th/9607201
\end{enumerate}

\vspace{5mm}

\section*{Acknowledgements}

I am grateful to prof. Tom Banks and to prof. Edward Witten for
valuable discussions especially on the non-scientifical issues
of my papers.

\end{document}